\documentclass[twocolumn]{revtex4}

\usepackage{graphicx}
\usepackage{amsfonts,amsmath,amssymb}

\newcommand{\moy}[1]{\langle #1 \rangle}
\newcommand{\moyqs}[1]{\langle #1 \rangle_\text{qs}}
\newcommand{\Moy}[1]{\Big\langle #1 \Big\rangle}

\begin{document}

\title{The survival probability of a branching random walk in presence of an
absorbing wall}
%\shorttitle{Survival probability of a BRW with an absorbing wall}

\author{Bernard Derrida}\email{bernard.derrida@lps.ens.fr}
\affiliation{Laboratoire de Physique Statistique,\\ \'Ecole
Normale Sup\'erieure,\\24, rue Lhomond, 75231 Paris Cedex 05,
France}

\author{Damien Simon}
\email{damien.simon@lps.ens.fr} 
\affiliation{Laboratoire de Physique Statistique,\\ \'Ecole Normale
Sup\'erieure,\\24, rue
Lhomond, 75231 Paris Cedex 05, France}

%\pacs{02.50.-r}{Probability theory, stochastic processes, and statistics}
%\pacs{05.40.-a}{Fluctuation phenomena, random processes, noise, and Brownian
%motion}
%\pacs{05.70.Jk}{Critical point phenomena}

\begin{abstract}
A branching random walk in presence
of an absorbing wall moving at a constant velocity $v$ undergoes a phase
transition as $v$ varies. The problem can be
analyzed using the properties of the Fisher-Kolmogorov-Petrovsky-Piscounov
(F-KPP) equation. We find that the survival probability of the branching random
walk vanishes at a critical velocity $v_c$ of the wall with an essential
singularity and we characterize the divergences of the relaxation times for
$v<v_c$ and $v>v_c$. At $v=v_c$ the survival probability decays like a
stretched exponential. Using the F-KPP equation, one can also calculate the
distribution of the population size at time $t$ conditionned by the survival of
one individual at a later time $T>t$. Our numerical results indicate that the
size of the population diverges like the exponential of $(v_c-v)^{-1/2}$ in the
quasi-stationary regime below $v_c$. Moreover for $v>v_c$, our data
indicate that there is no quasi-stationary regime.
\end{abstract}

\maketitle

There has been a long-standing interest in problems of non-equilibrium
critical phenomena and phase transition into absorbing states \cite{hinrichsen,
odor} (directed percolation or contact processes
\cite{dickman, domanykinzel},
reaction diffusion problems \cite{redner, cardytauber}). The goal of the
present letter is to analyse one such problem, which has been already considered
by several authors in the mathematical literature
\cite{bramson1,ferrari95,harrisharris}~: a branching random walk in presence of
an absorbing moving wall at a constant velocity $v$. For such a problem, there
is an obvious absorbing state (all the particles are absorbed by the wall) and
in this letter we analyse the critical behaviour of the survival
probability as the velocity $v$ of the wall varies. Our approach is based on an
analysis of the Fisher-Kolmogorov-Petrovsky-Piscounov (F-KPP) equation
\cite{fisher,kpp,bramson2, mueller, vansaarloos, panjasaarloos, brunetcutoff}
which is known to be related to branching random walks \cite{mckean, bramson1}.
We also show that the F-KPP equation allows us to calculate the properties of
the quasi-stationary regime.

In absence of an absorbing wall, a one dimensional branching 
random walk has a number of descendants which grows exponentially with time
 and these descendants occupy a region which spreads \cite{bramson1,mckean} in
space at
a known velocity $v_c$ (which can be calculated from the knowledge of the
branching and hopping rates of the walk).
 In presence of an absorbing wall moving at a constant velocity $v$, 
there is a competition between this exponential growth of the number
 of descendants and the absorption by the wall. 
One quantity of interest is the survival probability $Q(x,t)$, for a
particle starting at $t=0$ at distance $x$ from the wall, to have at
least one surviving descendant at time $t$. If $v> v_c$
(i.e. if the wall moves faster than the spreading velocity of the
population in absence of the wall), 
$Q(x,t) \to 0$ as $t\to \infty$. 
In this letter we study the phase transition which takes place at $v=v_c$ when
$v$ varies
\cite{harrisharris}. We will show that for $v < v_c$, the survival
 probability $Q(x,t)$ has a non-zero long time limit which vanishes with an
essential
 singularity as $v \to v_c$. We will also show that the characteristic time 
scale $\tau$ (on which $Q(x,t)$ converges exponentially to $Q(x,\infty)$)
diverges like $|v-v_c|^{-\frac{3}{2}}$ below $v_c$ and like $|v-v_c|^{-1}$ 
above $v_c$. At $v=v_c$, $Q(x,t)$ decays like a stretched exponential as
already proved by Kesten \cite{kesten}.

\begin{figure}
\begin{center}
\begin{picture}(0,0)%
\includegraphics[height=5cm]{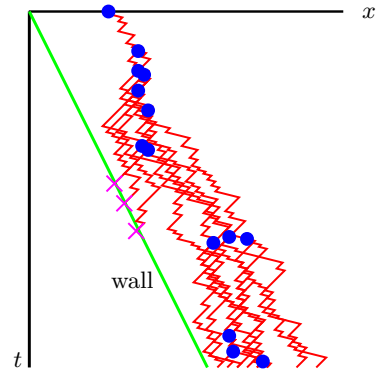}%
\end{picture}%
\begingroup
\setlength{\unitlength}{1cm}%
\begin{picture}(8,5)(0,0)%
\put(1.3,0.2){\makebox(0,0)[r]{\strut{} $t$}}%
\put(6,4.85){\makebox(0,0)[r]{\strut{} $x$}}%
\put(3.05,1.25){\makebox(0,0)[r]{\strut{} wall}}%
\end{picture}
\endgroup%
\end{center}
\caption{Branching random walk in presence of an absorbing wall. The circles
represent branching events and the crosses absorption by the wall.}
\end{figure}

Our motivation for the problem in presence of a moving
absorbing wall comes from the interest for simple models of
evolution
\cite{kloster,brunetgenea} proposed recently to study the \emph{in
vitro} evolution of a population of DNA molecules selected according to their
ability to bind to some target proteins \cite{dubertret}. In these models, the
position of an individual
represents its  adequacy to the environment (or fitness or ability to bind),
meaning that
when selection acts, the individuals with lowest $x$ are eliminated first.
The branching random walk has the effect of modeling the
growth of the population and the mutations through diffusion along the fitness
axis. In the models studied in \cite{penggerland,snyder,klostertang}, the size
of the population is kept fixed~: selection means that whenever new individuals
appear in the population, the ones with lowest $x$ are eliminated in order to
keep the size of the population constant. Here the role of the wall moving at
constant velocity is also to eliminate the less fitted individuals. Thus the
size
of the population can vary and even
vanish. Besides, the models of evolution studied in \cite{kloster,brunetgenea}
have a
broader interest in physics as they are also related to models that appear in
the theory of disordered systems \cite{derridaspohn},
in reaction-diffusion problems \cite{pecheniklevine,doeringmueller} or in
particle physics \cite{peschanski,iancumuellermunier,marquetpeschanski}.

In the first part of this letter, we consider a continuous branching random walk
in presence of a wall moving at constant velocity $v$ (for the
simulations below we will use a discrete space-time version to
facilitate the programming). The diffusion is described as follows~: during
$dt$, the position of an individual is shifted by an amount $\eta \sqrt{dt}$
where $\eta$ is a random variable of zero mean and of variance $2$.
For an initial individual at distance $x>0$ from the wall at $t=0$, one defines
the survival probability $Q(x,t)$ that at least one of its descendants is still
living at $t$ (i.e. $1-Q(x,t)$ is the probability that all the descendance of
the individual initially at position $x$ ahead of the wall has been absorbed by
the wall).
To write the evolution of the probability
$Q(x,t)$, it is convenient to decompose the time interval $[0,t+dt]$ into the
first time interval $dt$ and a remaining time interval $[dt,t+dt]$. During the
first interval $dt$, while the wall moves by $vdt$, the
single particle initially at position $x$ diffuses and can
branch with probability $\beta dt$ into two particles, as expressed by the two
terms of the following equation~:
\begin{equation}
\label{eq:fkppderiv}
\begin{split}
Q(x+ vdt,t+dt) =& \int \frac{  e^{-\eta^2/4} d\eta}{\sqrt{4\pi}} Q(x+ \eta
\sqrt{dt},t) \\ 
&+ \beta dt (Q(x,t) - Q(x,t)^2 )
\end{split}
\end{equation}
which becomes by taking the limit $dt \to 0$
\begin{equation}
 \label{eq:FKPP}
 \partial_t Q = \partial_{x}^2 Q - v\partial_x Q + \beta (Q-Q^2)  
\end{equation}
with the boundary conditions $Q(0,t)=0$ and $Q(x,0)=1$. It is important to
notice that (\ref{eq:FKPP}) is exact. The origin of the term $Q(x,t)^2$ is
that after a branching event, the two offspring have independent evolutions and
therefore their survival probabilities are uncorrelated.

All the $x$ and
$t$ dependence  of $Q(x,t)$ can be extracted from the analysis of
(\ref{eq:FKPP}) which is, up to the boundary conditions, the Fisher-KPP
 equation \cite{fisher,kpp}
in a moving frame at velocity $v$.

$Q(x,t)$ is obviously a monotonic function (increasing with $x$ and
decreasing with $t$~: it increases if one starts
further from the wall and it can only decrease with time). Therefore its $t \to
\infty$ limit $Q^*(x)$ exists
and satisfies
\begin{equation}
 \label{eq:FKPP-bis}
 \partial_{x}^2 Q^* - v\partial_x Q^* + \beta (Q^*-{Q^*}^2) =0 
\end{equation}
which is the usual equation for the shape of a front solution of the
F-KPP equation moving at velocity $v$ on the infinite line \cite{vansaarloos}. 

On the infinite line \cite{vansaarloos}, for every velocity $v$,
there is (up to a translation) a solution
$Q_v(x)$ of
(\ref{eq:FKPP-bis}) such that $Q_v(x) \to 0$ as $x \to -\infty$ and
$Q_v(x)=1$ as $x \to \infty$. These solutions are monotonic functions of
$x$ for $v > v_c \equiv 2 \sqrt{\beta}$ whereas  for $v < v_c$ they have damped
oscillations $ Q_v(x) \simeq A e^{\frac{v}{2} x} \cos [\sqrt{v_c^2 -
v^2}(x-x_0)/2] $ as $x \to
- \infty$. The relation between the decay of $Q_v(x)$ as
$x \to -\infty$ and the velocity is easy to obtain: one assumes that
$Q_v(x) \sim e^{\gamma x}$ in the region where $Q_v(x) \ll 1 $ and
one gets from (\ref{eq:FKPP-bis}) that  $v$ is related to $\gamma$ by
\begin{equation}
v(\gamma) = \gamma + \frac{\beta}{\gamma}.
\label{vgamma}
\end{equation}
The monotonic decay for $v> v_c$ or the damped
oscillations  $v<v_c$ are simply due to the fact that the solutions
$\gamma$ of the equation $v(\gamma)=v$ are real or complex conjugate
$\gamma=\gamma_R\pm i\gamma_I$. It is also known that, at $v=v_c$, the
asymptotic decay has the form
\begin{equation}
Q_{v_c}(x) \simeq (- A  (x-x_0) + B)  e^{\gamma_c (x-x_0) }
\label{Qvc}
\end{equation}
where $x_0$ is arbitrary and $\gamma_c= \sqrt{\beta}$ is the value of $\gamma$
for which $v(\gamma)
$ is minimum (and where $v(\gamma_c)= v_c$) .

\begin{table*}[!ht]
\begin{center}
 \begin{tabular}{|c|c|c|c|}
  \hline
Velocity	& $v\lesssim v_c$	& $v=v_c$	& $v \gtrsim v_c$ \\
\hline
 $Q^*(x)$	&  $\sim
\exp\Big(-\gamma_c\sqrt{\frac{\pi^2v''(\gamma_c)}{2(v_c-v)}}\Big)$	& 0	
& 0 \\
\hline
Relaxation & $\sim e^{-t/\tau}$	& $\sim e^{ -\gamma_c \big( 3\pi^2
v''(\gamma_c)/2\big)^{1/3} t^{1/3}} $&  $\sim e^{-t/\tilde{\tau}}$ \\
\hline
Times $\tau$, $\tilde{\tau}$ &
$\tau\simeq \frac{1}{2}\sqrt{\frac{\pi^2
v''(\gamma_c)}{2}}(v_c-v)^{-\frac{3}{2}}$ &   
&$\tilde{\tau}\simeq\frac{1}{\gamma_c}
(v-v_c)^{-1}$
 \\
\hline
 \end{tabular}
\end{center}
\caption{\label{tab:tabcritical} Critical behaviour
(\ref{eq:Q:essentialsing},\ref{eq:tau},\ref{eq:Qstretched},\ref{eq:Q:above})
near $v_c$ of the
survival probability $Q(x,t)$ at large time $t$ for an arbitrary $v(\gamma)$.}
\end{table*}

With the boundary condition $
Q^*(0)=0$
the solution of (\ref{eq:FKPP-bis}) 
 is obviously $Q^*(x)=0$ for $v > v_c$ as the wall moves faster 
than the spreading velocity $v_c$ of the population.
On the other hand, for $v< v_c$, the solution is the right-most
positive arch of the solution on the infinite line.
As $v$ approaches $v_c$, the region where $Q^*$ is small and where
the quadratic term in (\ref{eq:FKPP-bis}) can be neglected becomes larger
and larger.
For $0 < v_c-v \ll 1 $ the solution $Q^*$ can be
decomposed into two parts: a region  $ x > L$ where  $Q^*(x)$ is of order
$1$ and resembles the solution $Q_{v_c}(x)$ on the
infinite line for $v=v_c$ (whose asymptotics decay is (\ref{Qvc})) and a region
$0<x<L$ where $Q^*(x)$ is small and
of the form $Q^*(x)=C\sin (\gamma_I x) e^{\gamma_R x}$. To match the asymptotic
decay of (\ref{Qvc}) near $x=L$, one should take $L=\pi/\gamma_I=2 \pi /
\sqrt{v_c^2 -
v^2}$ and $C=Ae^{-\gamma_R L}/\gamma_I$ such that
\begin{equation}
Q^*(x) \simeq A L \sin \left( \frac{\pi x}{L} \right) e^{\gamma_R (x - L)
}.
\label{Qlinear}
\end{equation}
This implies that as $v \to v_c^-$, the survival probability vanishes with
an essential singularity
\begin{equation}
\label{eq:Q:essentialsing}
Q^*(x) \sim \exp \left[  - \frac{\pi \sqrt{\gamma_c} }{
\sqrt{v_c-v}} +
\gamma_c x  \right].
\end{equation}
One can remark that a shape of $Q^*(x)$ made up of two parts is almost exactly
the
same as the shape of a moving front on the infinite line, in presence of a
cut-off \cite{brunetcutoff}.

To obtain the long time dependence of $Q(x,t)$ we assume that one can still
decompose  $Q(x,t)$ into two parts~: a region
$0<x<L_t$ where $Q(x,t)$ is small with a solution of the form
\begin{equation}
\label{eq:sinearch}
Q(x,t)= A L_t \sin\Big( \frac{\pi x}{L_t} \Big) e^{\gamma_c (x-L_t)}
\end{equation}
(this assumption will be tested in the figure \ref{fig:sinearch} below)
and a region $x> L_t$ where $Q(x,t)$ is not small and resembles (\ref{Qvc}).
One can check that (\ref{eq:sinearch}) is indeed solution of  
(\ref{eq:FKPP-bis})
in the range $0<x<L_t$ to leading non-vanishing order in $L_t^{-1}$ and $v-v_c$
provided that $L_t$ satisfies
\begin{equation}
\label{eq:Lt:evol}
\partial_t L_t = v-v_c +\frac{\pi^2}{\gamma_c L_t^2}.
\end{equation}
As one expects  (at least for $v<v_c$) that $L_t/L \sim L_t
(v_c-v)^{1/2} \to 0$ as 
$t \to 0$, this determines $L_t$ and one gets that for $0 < v_c-v \ll 1$,
\begin{equation}
\label{eq:Lt:scaling}
L_t=\frac{\pi}{\sqrt{\gamma_c (v_c-v)}} F\left( \frac{\sqrt{\gamma_c}}{\pi}
(v_c-v)^{3/2} t \right)
\end{equation}
where the function $F(z)$ satisfies $\partial_z F = -1+ F^{-2}$ whose solution 
 (given that $F(0)=0$) is 
\begin{equation}
\label{eq:Fsol}
 -2 F + \ln \frac{1+F}{ 1-F} = 2 z \ \ \ \  \text{i.e.} \ \ \ \ F =
\tanh(F+z).
\end{equation}
For large $z$, one gets  $F\simeq 1 - 2 e^{-2 z - 2}$. Therefore
(\ref{eq:Lt:scaling}) implies that for $v\to v_c^-$ the convergence time
$\tau$ of the exponential decay of $Q(x,t)$
 towards the fixed profile $Q^*(x)$ is
\begin{equation}
\label{eq:tau}
 \tau \sim (v_c-v)^{-3/2} \pi/(2 \sqrt{\gamma_c})
\end{equation}
For small $z$, one can expand the solution of (\ref{eq:Fsol})
$F(z) = (3z)^{1/3} - 3 z / 5 + O(z^{5/3})$.
This leads to  $L_t \simeq (3 \pi^2 /\gamma_c)^{1/3}  t^{1/3} + O(
(v-v_c) t)$ and therefore to a stretched exponential decay of $Q(x,t)$ at
$v=v_c$
\begin{equation}
\label{eq:Qstretched}
Q(x,t) \sim \exp[- (3 \pi^2/\gamma_c)^{1/3} t^{1/3} ] .
\end{equation}
At any finite $t$, one can notice (\ref{eq:Lt:scaling},\ref{eq:Fsol}) that $L_t$
can be expanded in powers of $v-v_c$ and one can get the whole time dependence
for $v> v_c$ by an analytic continuation of the case $v < v_c$. In particular,
the scaling function $\widetilde{F}$ which replaces (\ref{eq:Lt:scaling}) for
$v-v_c>0$
is related analytically to the function $F$ obtained for $v-v_c <0$ and
(\ref{eq:Lt:scaling},\ref{eq:Fsol}) are replaced by~:
\begin{equation}
\begin{split}
 L_t=&\frac{\pi}{\sqrt{\gamma_c (v-v_c)}} \widetilde{F}\left(
\frac{\sqrt{\gamma_c}}{\pi}
(v-v_c)^{3/2} t \right) \\
 \widetilde{F}(z) =& \tan( \widetilde{F}(z) -z).
\end{split}
\end{equation}
Clearly $\widetilde{F}(z)=z+\pi/2-1/z + O(1/z^2)$ for large $z$. Thus for
$0<v-v_c \ll 1$ and $(v-v_c)^{3/2} t \gg 1$, one has $L_t \simeq (v-v_c)t$ and
for fixed $x$~:
\begin{equation}
\label{eq:Q:above}
Q(x,t) \sim \exp[\gamma_c (x-(v-v_c)t)].
\end{equation}
Both (\ref{eq:Qstretched}) and (\ref{eq:Q:above}) agree with earlier results by
Kesten \cite{kesten} and Harris and Harris \cite{harrisharris}.

\bigskip
To study numerically the time evolution of the survival probability $Q(x,t)$, it
is more convenient to use a discrete space time version of the
problem. Here we consider the case where $x$ takes only integer values. At
every time step, $t \to t+1$,
each individual has $k$ offspring and each offspring has a probability
$p_y$ of being produced at position $x + y$. 
Then the time evolution of $Q(x,t)$ which generalizes (\ref{eq:FKPP})  is
given by
\begin{equation}\label{eq:evol:discrete}
Q(x+v,t+1) = 1 - \left( 1 - \sum_y p_y Q(x+y,t) \right)^k
\end{equation}
and by looking at the region where  $Q(x,t) \sim e^{\gamma x }$ is small, the
expression (\ref{vgamma}) of the velocity $v(\gamma)$  becomes
\begin{equation*}
v(\gamma) = \frac{1}{\gamma} \log \left[ k \sum_y p_y e^{\gamma
y} \right].
\end{equation*}
As before, the spreading velocity $v_c$ of the population in absence of the
wall is $v_c = \min_\gamma v(\gamma)$.
In this case one can generalize the above calculations to an arbitrary
$v(\gamma)$ and the results
(\ref{eq:Q:essentialsing},\ref{eq:tau},\ref{eq:Qstretched},\ref{eq:Q:above})
are replaced by the expression of table \ref{tab:tabcritical}.

\begin{figure*}
 \begin{center}
%GNUPLOT: LaTeX picture with Postscript
\begin{picture}(0,0)%
\includegraphics{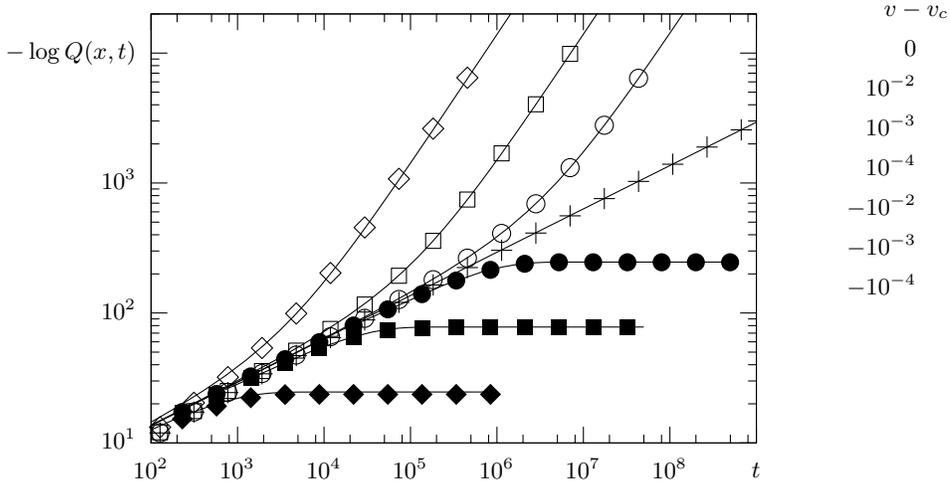}%
\end{picture}%
\begingroup
\setlength{\unitlength}{0.0200bp}%
\begin{picture}(19980,8640)(0,0)%
\put(2750,0){\makebox(0,0)[r]{\strut{}$10^{1}$}}%
\put(2750,2451){\makebox(0,0)[r]{\strut{}$10^{2}$}}%
\put(2750,4902){\makebox(0,0)[r]{\strut{}$10^{3}$}}%
\put(2750,7352){\makebox(0,0)[r]{\strut{}$-\log Q(x,t)$}}%
\put(3025,-550){\makebox(0,0){\strut{}$10^{2}$}}%
\put(4654,-550){\makebox(0,0){\strut{}$10^{3}$}}%
\put(6284,-550){\makebox(0,0){\strut{}$10^{4}$}}%
\put(7913,-550){\makebox(0,0){\strut{}$10^{5}$}}%
\put(9542,-550){\makebox(0,0){\strut{}$10^{6}$}}%
\put(11171,-550){\makebox(0,0){\strut{}$10^{7}$}}%
\put(12801,-550){\makebox(0,0){\strut{}$10^{8}$}}%
\put(14430,-550){\makebox(0,0){\strut{}$t$}}%
\put(17455,8165){\makebox(0,0){\strut{}$ v-v_c$}}%
\put(17455,7415){\makebox(0,0)[r]{\strut{}$ 0$}}%
\put(17455,6665){\makebox(0,0)[r]{\strut{}$ 10^{-2}$}}%
\put(17455,5915){\makebox(0,0)[r]{\strut{}$ 10^{-3}$}}%
\put(17455,5165){\makebox(0,0)[r]{\strut{}$ 10^{-4}$}}%
\put(17455,4415){\makebox(0,0)[r]{\strut{}$ -10^{-2}$}}%
\put(17455,3665){\makebox(0,0)[r]{\strut{}$ -10^{-3}$}}%
\put(17455,2915){\makebox(0,0)[r]{\strut{}$-10^{-4}$}}%
\end{picture}%
\endgroup
%\endinput 
 \end{center}
\caption{\label{fig:Qtvc+} The points represent the logarithm of the survival
probability $Q(x,t)$ at
$x=1$ for different values of $v-v_c$ both below and above the critical
velocity, obtained by iterating (\ref{eq:evol:discrete}). The lines are
the theoretical predictions (\ref{eq:sinearch},\ref{eq:Lt:scaling}). }
\end{figure*}

To facilitate the analysis and avoid
periodic or aperiodic dependence due to the commensurability of $v$ we decided
to fix $v=1$ for the velocity of the wall and to vary $v_c$ by changing the
$p_y$. In all the figures below, we show
the case where $p_2= 1/18 + \delta, p_1=1/18, p_0= 16/18- \delta$. Then by
varying $\delta$ we can vary $v_c$  and therefore the difference $v_c- v$ (which
vanishes for $\delta=0$). More precisely, for small $v_c-v$, one has the
first-order relation $\delta\simeq 15 (v_c-v)/(2\ln{4}) $ in the present case.

\begin{figure}
\begin{center}
%GNUPLOT: LaTeX picture with Postscript
\begin{picture}(0,0)%
\includegraphics{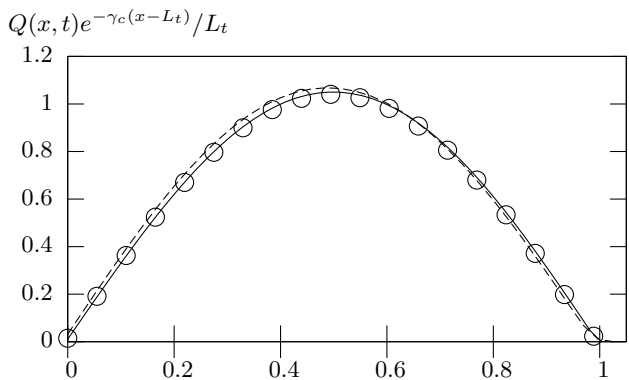}%
\end{picture}%
\begingroup
\setlength{\unitlength}{0.0200bp}%
\begin{picture}(10800,6480)(0,0)%
\put(0,0){\makebox(0,0)[r]{\strut{} 0}}%
\put(0,897){\makebox(0,0)[r]{\strut{} 0.2}}%
\put(0,1793){\makebox(0,0)[r]{\strut{} 0.4}}%
\put(0,2690){\makebox(0,0)[r]{\strut{} 0.6}}%
\put(0,3587){\makebox(0,0)[r]{\strut{} 0.8}}%
\put(0,4483){\makebox(0,0)[r]{\strut{} 1}}%
\put(0,5380){\makebox(0,0)[r]{\strut{} 1.2}}%
\put(-1000,6000){\makebox(0,0)[l]{\strut{} $Q(x,t)e^{-\gamma_c(x-L_t)}/L_t$}}%
\put(275,-550){\makebox(0,0){\strut{} 0}}%
\put(2280,-550){\makebox(0,0){\strut{} 0.2}}%
\put(4285,-550){\makebox(0,0){\strut{} 0.4}}%
\put(6289,-550){\makebox(0,0){\strut{} 0.6}}%
\put(8294,-550){\makebox(0,0){\strut{} 0.8}}%
\put(10299,-550){\makebox(0,0){\strut{} 1}}%
\end{picture}%
\endgroup
%\endinput
\end{center}
\caption{\label{fig:sinearch} Shape of the rescaled survival probability
$Q(x,t)e^{-\gamma_c(x-L_t)}/L_t$ in the linear domain at different times
$t=10^5$ (dashed), $5.10^5$ (circle), $11.10^5$ (line). The corresponding
lengths are $L_t=1.05.10^2, 2.01.10^2, 2.93.10^2$) for $v-v_c\simeq 10^{-4}>0$
($\delta= \frac{15}{2\ln{4}} (v_c-v)$ at first order for this particular model).
The length $L_t$ is measured directly as the point where $Q(L_t,t)=0.5$ by
linear
interpolation between the sites of the lattice. The observed shape agrees with
(\ref{eq:sinearch}).}
\end{figure}

In figure \ref{fig:sinearch} we test our assumption (\ref{eq:sinearch}). For
an initial individual at $x=1$, i.e. for $Q(x,0)=\delta_{x,1}$, we computed
$Q(x,t)$ by iterating (\ref{eq:evol:discrete}). We plot
$Q(x,t)e^{-\gamma(x-L_t)}/L_t$ which according to (\ref{eq:sinearch}) should
depend on $x/L_t$ only. In figure \ref{fig:Qtvc+}, we compare the values of
$Q(1,t)$ obtained by iterating (\ref{eq:evol:discrete}) with the theoretical
predictions (\ref{eq:sinearch}, \ref{eq:Lt:scaling}). The agreement is
excellent.

\bigskip

At $t=\infty$, the size of the population is either zero
(with probability $1$ if $v\geq v_c$ and $1-Q(x,\infty)$ if $v<v_c$) or
infinite. At large but finite final time $T$, there are however events
corresponding, for example, to having exactly one survivor. One can then try to
understand
which strategy leads to such events. For $v<v_c$, $t\gg 1$ and $T-t=t' \gg 1$,
the system adopts a quasi-stationary regime during which its
properties do not depend on the initial conditions (number and positions of the
individuals at $t=0$). In particular, the average population size at time $t$
conditioned by the survival of only one individual at final time $T=t+t'$, do
not depend on $t$ as long as $t \gg 1$ and $t'\gg 1$.

We are now going to sketch how the properties of this quasistationnary regime
can be computed from (\ref{eq:FKPP}) or (\ref{eq:evol:discrete}). Details
will be published in a future work \cite{derridasimon}. Let
us introduce the generating function~:
\begin{equation}
 \label{eq:defG}
 G(x,t ; f) = \Moy{ \prod_{i=1}^{N_t} e^{-f( x_i^{(t)} )} }
\end{equation}
where $f$ is an arbitrary function, $N_t$ is the size of the population at
time $t$ and $x_1^{(t)},\ldots,x_{N_t}^{(t)}$  are the positions of individuals
at time $t$. By analysing what happens during the very first instant $d\tau$,
as in (\ref{eq:fkppderiv}), one finds that~:
\begin{equation}
\label{eq:G:rec}
\begin{split}
 G( x+vdt , t+dt ; f) =& \beta dt
\Big(G(x,t;f)^2-G(x,t ;f)\Big)
\\
& + \int \frac{  e^{-\eta^2/4} d\eta}{\sqrt{4\pi}} G(x+ \eta\sqrt{dt},t;f).
\end{split}
\end{equation}
Therefore $1-G(x,t ;f)$ satisfies the F-KPP equation (\ref{eq:FKPP}) with the
boundary conditions $G(x=0, t ;f)=1$ and $G(x, t=0 ;f)=e^{-f(x)}$. In the
discrete space time case, a similar reasoning  shows that $1-G$ satisfies
(\ref{eq:evol:discrete}).

One can also consider generating functions at two (or more) different times
$0<t<t+t'$ defined as~:
\begin{equation}
 H(x, t, t' ; f_1, f_2) = \Moy{  \prod_{i=1}^{N_{t}} e^{-f_1(
x_i^{(t)} )} \prod_{j=1}^{N_{t+t'}} e^{-f_2( x_j^{(t+t')} )}}.
\end{equation}
Like in (\ref{eq:G:rec}), one can show that $1-H(x, t, t' ; f_1, f_2)$ as a
function of $x$ and $t$ satisfies (\ref{eq:fkppderiv}) or
(\ref{eq:evol:discrete}) with boundary conditions $H(x=0,t,t' ; f_1,f_2)=1$ and
$H(x,0,t'; f_1, f_2) = e^{-f_1(x)} G(x,t' ; f_2)$ which, as we have just seen,
can also be determined from (\ref{eq:fkppderiv}) or (\ref{eq:evol:discrete}).

If one chooses $f_1=\lambda$ and $f_2=\mu$, then $H(x,t,t';f_1,f_2)= \moy{
e^{-\lambda N_t -\mu N_{t+t'}}}$ is the generating function of the sizes of the
population at two different times. If one expands $H$ to first order
in powers of $e^{-\mu}$ and of $\lambda$
\begin{equation}
 \label{eq:expansionH}
\begin{split}
 H(x,t,t' ; \lambda ,\mu) \simeq& H_{00}(x,t,t') + e^{-\mu}H_{10}(x,t,t') + \\
&\lambda H_{01}(x,t,t') + \lambda e^{-\mu} H_{11}(x,t,t') 
\end{split}
\end{equation}
one gets that the average size of the population $\moy{N_t}_{T=t+t'}$ at time
$t$ conditionned on having a single survivor left at time
$T=t+t'$ is given by~:
\begin{equation}
\label{eq:Nmoy:H}
\moy{N_t}_{T=t+t'}=-H_{11}(x,t,t')/H_{10}(x,t,t').
\end{equation}

\begin{figure*}[!ht]
 \begin{center} 
  %GNUPLOT: LaTeX picture with Postscript
\begin{picture}(0,0)%
\includegraphics{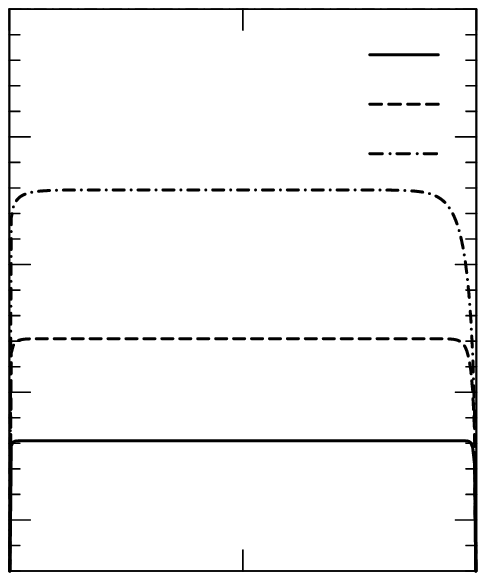}%
\end{picture}%
\begingroup
\setlength{\unitlength}{0.0200bp}%
\begin{picture}(8099,8640)(0,0)%
\put(825,735){\makebox(0,0)[r]{\strut{}$10^{5}$}}%
\put(825,2574){\makebox(0,0)[r]{\strut{}$10^{10}$}}%
\put(825,4413){\makebox(0,0)[r]{\strut{}$10^{15}$}}%
\put(825,6251){\makebox(0,0)[r]{\strut{}$10^{20}$}}%
\put(825,8090){\makebox(0,0)[r]{\strut{}$\moy{N_t}_T$}}%
\put(7825,-550){\makebox(0,0){\strut{} $t$}}%
\put(4462,-550){\makebox(0,0){\strut{} 125000}}%
\put(1100,-550){\makebox(0,0){\strut{} 0}}%
\put(6016,7434){\makebox(0,0)[r]{\strut{}$v-v_c=-10^{-2}$}}%
\put(6016,6722){\makebox(0,0)[r]{\strut{}$-5.10^{-3}$}}%
\put(6016,6010){\makebox(0,0)[r]{\strut{}$-10^{-3}$}}%
\end{picture}%
\endgroup
 \hfil
%GNUPLOT: LaTeX picture with Postscript
\begin{picture}(0,0)%
\includegraphics{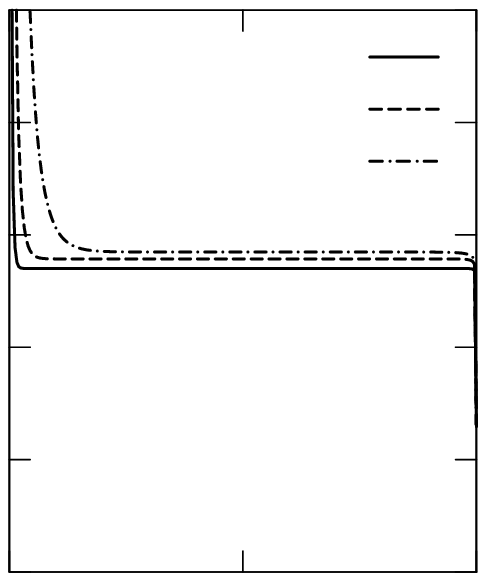}%
\end{picture}%
\begingroup
\setlength{\unitlength}{0.0200bp}%
\begin{picture}(8099,8640)(0,0)%
\put(825,0){\makebox(0,0)[r]{\strut{} 0.5}}%
\put(825,1618){\makebox(0,0)[r]{\strut{} 1}}%
\put(825,3236){\makebox(0,0)[r]{\strut{} 1.5}}%
\put(825,4854){\makebox(0,0)[r]{\strut{} 2}}%
\put(825,6472){\makebox(0,0)[r]{\strut{} 2.5}}%
\put(825,8090){\makebox(0,0)[r]{\strut{}
$\dfrac{\moy{N_t^2}_T}{\moy{N_t}_T^2}$}}%
\put(7825,-550){\makebox(0,0){\strut{} $t$}}%
\put(4462,-550){\makebox(0,0){\strut{} 125000}}%
\put(1100,-550){\makebox(0,0){\strut{} 0}}%
\put(6016,7415){\makebox(0,0)[r]{\strut{}$v-v_c=-10^{-2}$}}%
\put(6016,6665){\makebox(0,0)[r]{\strut{}$-5.10^{-3}$}}%
\put(6016,5915){\makebox(0,0)[r]{\strut{}$-10^{-3}$}}%
\end{picture}%
\endgroup
\\
 \end{center}
 \caption{\label{fig:quasistatN} On the left side~: average size of the
population at $0\leq t \leq T$ conditionned by the existence of a unique
survivor at $T=250000$ for different values of $v-v_c < 0$. The results
are obtained using (\ref{eq:evol:discrete},\ref{eq:expansionH},\ref{eq:Nmoy:H}).
On the right side, the ratio
$\moy{N_t^2}_T/\moy{N_t}_T^2$ is represented for $0\leq t \leq T$ for the same
values of $v-v_c<0$. In both cases, the initial individual at $t=0$ is at
$x=1$. A quasi-stationary regime (with constant $\moy{N_t}_T$ and
$\moy{N_t^2}_T/\moy{N_t}_T^2$) is observed
as soon as $T$ is much larger than the relaxation time $\tau$.}
\end{figure*}

Numerical values obtained by iterating the equations satisfied by
$H_{00}$, $H_{10}$, $H_{01}$ and $H_{11}$ (derived by replacing $H$ by its
expansion (\ref{eq:expansionH}) into (\ref{eq:evol:discrete})) are shown in
figure (\ref{fig:quasistatN}). By pushing further the expansion
(\ref{eq:expansionH}) in powers of $\lambda$, one can have access to higher
moments of the population size $N_t$.

In figure \ref{fig:quasistatN}, we see that for $v<v_c$ both $\moy{N_t}_T$ and $
\moy{N_t^2}_T/\moy{N_t}_T^2$ become constant over a very long time interval,
indicating that there is a quasi-stationary regime. As $v\to v_c$, the size of
the population in this quasi-stationary regime diverges and the ratio $
\moy{N_t^2}_T/\moy{N_t}_T^2$ seems to converge to a value close to $2$. This
indicates that the fluctuations of the population size $N_t$ in the
quasi-stationary regime are of the same order as $\moy{N_t}$ itself.

For $v<v_c$ in fact, the whole distribution of $N_t$ at $0\ll{}t$ conditionned
on the survival of a single individual at $T=t+t'>t$ is
independent of the position $x$ of the initial individual and reaches a
quasi-stationary regime \cite{ferrari95} before a final relaxation to the final
state containing
only one individual. In this quasi-stationary regime one can show from
(\ref{eq:FKPP}) and (\ref{eq:expansionH},\ref{eq:Nmoy:H}) that the average size
of the population $\moyqs{N_t}$ can be calculated for $v\to v_c$ (details will
be published in \cite{derridasimon}) and one finds
\begin{equation}
\label{eq:Nqs:div}
 \moyqs{N_t}\sim \exp\left[ \gamma_c \sqrt{\frac{\pi^2 v''(\gamma_c)}{2
(v_c-v)}}\right].
\end{equation}

\begin{figure}
 \begin{center}
\begin{picture}(0,0)%
\includegraphics{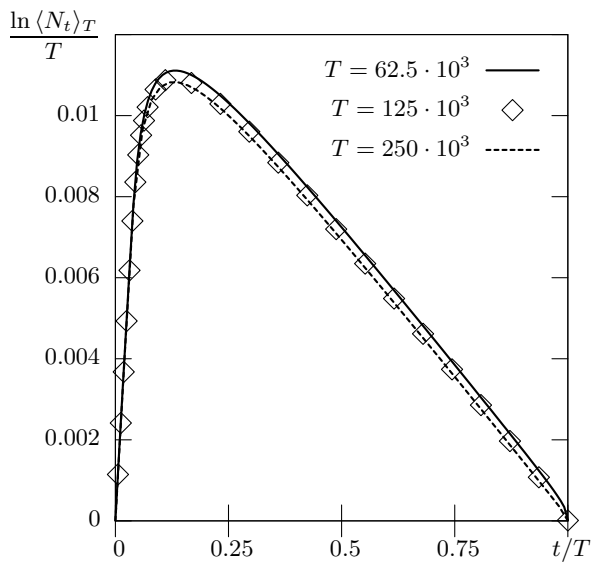}%
\end{picture}%
\begingroup
\setlength{\unitlength}{0.0200bp}%
\begin{picture}(9900,9719)(0,0)%
\put(825,0){\makebox(0,0)[r]{\strut{} 0}}%
\put(825,1528){\makebox(0,0)[r]{\strut{} 0.002}}%
\put(825,3057){\makebox(0,0)[r]{\strut{} 0.004}}%
\put(825,4585){\makebox(0,0)[r]{\strut{} 0.006}}%
\put(825,6113){\makebox(0,0)[r]{\strut{} 0.008}}%
\put(825,7642){\makebox(0,0)[r]{\strut{} 0.01}}%
\put(825,9170){\makebox(0,0)[r]{\strut{} $\dfrac{\ln{\moy{N_t}_T}}{T} $}}%
\put(9625,-550){\makebox(0,0){\strut{} $t/T$}}%
\put(7494,-550){\makebox(0,0){\strut{} 0.75}}%
\put(5362,-550){\makebox(0,0){\strut{} 0.5}}%
\put(3231,-550){\makebox(0,0){\strut{} 0.25}}%
\put(1100,-550){\makebox(0,0){\strut{} 0}}%
\put(7816,8495){\makebox(0,0)[r]{\strut{}$T=62.5\cdot 10^3$}}%
\put(7816,7745){\makebox(0,0)[r]{\strut{}$T=125\cdot 10^3$}}%
\put(7816,6995){\makebox(0,0)[r]{\strut{}$T=250\cdot 10^3$}}%
\end{picture}%
\endgroup
%\endinput
 \end{center}
\caption{\label{fig:noquasistat} The logarithm of the average size of the
population divided by $T$ at $0\leq t \leq T$ conditionned by the existence of a
unique survivor at $T$ is plotted as a function of the rescaled time $t/T$ for
$v-v_c=10^{-2}$ and different final times $T=62500,125000,250000$. Unlike
figure \ref{fig:quasistatN}, no quasi-stationary regime is observed.}
\end{figure}

For $v>v_c$, no quasi-stationary state is observed (see
figure~\ref{fig:noquasistat}). Instead, the evolution
of the system conditionned on having a single survivor at a later time
$T$ can be divided into two parts. The first one corresponds to a
rapid growth of the population, which can be described by the
modified dynamics introduced in \cite{harrisharris}, followed by a regular
absorption of individuals by the wall leading to the survival of exactly one
individual at $T$. Numerical results obtained from (\ref{eq:evol:discrete}) and
(\ref{eq:expansionH},\ref{eq:Nmoy:H}) are presented in figure
\ref{fig:noquasistat}~: they indicate that $\moy{N_t}_T$ takes a scaling form
$\moy{N_t}_T\sim \exp[Tf(t/T)]$.

\bigskip

In this article, we have seen how a branching random walk in presence of a
moving absorbing walk undergoes a phase transition at a critical velocity
$v_c$ with a survival probability which vanishes with an essential singularity
(\ref{eq:Q:essentialsing}) and decays with time like a stretched exponential
(\ref{eq:Qstretched}) at the transition. We have also seen how the properties of
the quasi-stationary regime \cite{ferrari95} can be computed
(\ref{eq:defG}--\ref{eq:Nmoy:H}) from the travelling wave equation.

It is interesting to notice that several of our results are very reminiscent of
what is known of noisy F-KPP like equation \cite{brunetcutoff,pecheniklevine}.
First the relation (\ref{eq:Nqs:div}) between the size $N$ of the population in
the quasi-stationary regime and the velocity can be rewritten as~:
\begin{equation}
 v-v_c\simeq -\frac{\pi^2 v''(\gamma_c)}{2\gamma_c^2}\frac{1}{\ln^2 N}
\end{equation}
which is exactly the prediction of the cut-off approximation
\cite{brunetcutoff, pecheniklevine,brunetpheno}. Moreover, for $v<v_c$, the
relaxation time $\tau\sim (v_c-v)^{-3/2}$ can be rewritten as~:
\begin{equation}
 \tau \sim \ln^3 N
\end{equation}
which is exactly what was observed recently for the coalescence times in the
genealogies of evolution models \cite{brunetgenea, brunetgenea2}
related to these noisy travelling wave equations.

Our results on the quasi-stationary regime presented here (eq.
(\ref{eq:Nqs:div}) and figures \ref{fig:quasistatN} and \ref{fig:noquasistat})
are preliminary and will be developed  in a forthcoming work
\cite{derridasimon}.

Beyond the understanding of this quasi-stationary regime, it would be
interesting to see whether our main results which are based on our guess
(\ref{eq:sinearch}) for the shape of the solution of (\ref{eq:FKPP}) could be
recovered by more standard methods like the renormalization group approach
\cite{cardytauber} or the phenomenological picture \cite{brunetpheno} of noisy
fronts. Among other possible extensions of the present work, one could try to
study how our results would be modified when the position of the wall has a
more complicated evolution than a constant velocity \cite{braysmith}.

\end{document}